\documentclass[usenatbib]{mn2e}
\usepackage{graphicx, txfonts, natbib}
\usepackage{epsf}
\usepackage{amssymb}
\usepackage{epstopdf}
\usepackage{longtable}
\usepackage{subfigure}
\newcommand{\bv}{\mbox{$B\!-\!V$}}

\newcommand{\msun}{\mbox{M$_{\odot}$}}
\newcommand{\msol}{\mbox{M$_{\odot}$}}

\newcommand{\nick}{\mbox{$^{56}$Ni}}
\newcommand{\cob}{\mbox{$^{56}$Co}}
\DeclareMathAlphabet{\mathsc}{OT1}{cmr}{m}{sc}
\def\testbx{bx}%
\DeclareRobustCommand{\ion}[2]{%
\relax\ifmmode
\ifx\testbx\f@series
{\mathbf{#1\,\mathsc{#2}}}\else
{\mathrm{#1\,\mathsc{#2}}}\fi
\else\textup{#1\,{\mdseries\textsc{#2}}}%
\fi}

\newcommand{\Nai}{\ion{Na}{i}}
\newcommand{\Feii} {\ion{Fe}{ii}}

\newcommand{\Caii} {\ion{Ca}{ii}}

\newcommand{\Cii} {\ion{C}{ii}}
\newcommand{\Crii} {\ion{Cr}{ii}}

\newcommand{\Oinir} {\ion{O}{i}}

\newcommand{\SiII} {\ion{Si}{ii}}

\newcommand{\Suii}{\ion{S}{ii}}

\newcommand{\Mgi} {\ion{Mg}{i}}
\newcommand{\Mgii} {\ion{Mg}{ii}}
\newcommand{\Scii} {\ion{Sc}{ii}}
\newcommand{\TIii} {\ion{Ti}{ii}}

\begin{document}
\title[PTF10ops: A subluminous but normal-width SN Ia]
  {PTF10ops -- a subluminous, normal-width lightcurve Type Ia supernova in the
middle of nowhere}

\author[K. Maguire et al.]
  {K.~Maguire$^1$\thanks{E-mail: kate.maguire@astro.ox.ac.uk}, 
 M.~Sullivan$^1$, R.~C.~Thomas$^{2,3}$, P.~E.~Nugent$^{2,3}$,  D.~A.~Howell$^{4,5}$, A.~Gal-Yam$^6$, 
  \newauthor  I.~Arcavi$^6$, S.~Ben-Ami$^6$, S.~Blake$^1$, J.~Botyanszki$^{2,7}$, C.~Buton$^8$, J.~Cooke$^{9}$,  R.~S.~Ellis$^{10}$,   
       \newauthor   I.~M.~Hook$^{1,11}$,  M.~M.~Kasliwal$^{10}$, Y.-C.~Pan$^1$, R.~Pereira$^{12}$,  P.~Podsiadlowski$^{1}$,   
   \newauthor  A.~Sternberg$^6$,  N.~Suzuki$^2$, D.~Xu$^6$, O.~Yaron$^6$, J.~S.~Bloom$^3$, S.~B.~Cenko$^3$,   S.~R.~Kulkarni$^{10}$,
   \newauthor  N.~Law$^{13}$, E.~O.~Ofek$^{10}$, D.~Poznanski$^{2,3}$, R.~M.~Quimby$^{10}$\\
     $^1$Department of Physics (Astrophysics), University of Oxford, DWB, Keble Road, Oxford OX1 3RH, UK\\
    $^2$Computational Cosmology Center, Lawrence Berkeley National Laboratory, 1 Cyclotron Rd., Berkeley CA 94720, USA\\
   $^3$Department of Astronomy, University of California, Berkeley, CA 94720-3411, USA \\
   $^4$Las Cumbres Observatory Global Telescope Network, 6740 Cortona Dr., Suite 102, Goleta, CA 93117, USA\\
   $^5$Department of Physics, University of California Santa Barbara, Santa Barbara, CA 93196, USA \\
   $^6$Department of Particle Physics and Astrophysics, Faculty of Physics, The Weizmann Institute of Science, Rehovot 76100, Israel\\
   $^7$Department of Physics, University of California, Berkeley, CA 94720-3411, USA\\
   $^8$Physikalisches Institut Universitat Bonn, Nussallee 12 53115 Bonn, Germany\\
   $^{9}$Swinburne University of Technology, Victoria 3122, Australias\\
   $^{10}$Cahill Center for Astrophysics, California Institute of Technology, Pasadena, CA 91125, USA\\
  $^{11}$INAF - Osservatorio Astronomico di Roma, via Frascati 33, 00040 Monteporzio (RM), Italy\\
   $^{12}$Universit\'e de Lyon, F-69622 France; Universit\'e de Lyon 1, Villeurbanne; CNRS/IN2P3, Institut de Physique Nucl\'eaire de Lyon, France \\
   $^{13}$Dunlap Institute for Astronomy and Astrophysics, University of Toronto, 50 St. George Street, Toronto M5S 3H4, Ontario, Canada\\
    }

\maketitle

\begin{abstract}

PTF10ops is a Type Ia supernova (SN Ia), whose lightcurve and spectral properties place it outside the current SN Ia subtype classifications. Its spectra display the characteristic lines of subluminous SNe Ia, but it has a normal-width lightcurve with a long rise-time, typical of normal luminosity SNe Ia. The early-time optical spectra of PTF10ops were modelled using a spectral fitting code and found to have all the lines typically seen in subluminous SNe Ia, without the need to invoke more uncommon elements. The host galaxy environment of PTF10ops is also unusual with no galaxy detected at the position of the SN down to an absolute limiting magnitude of $r\geq-12.0$ mag, but a very massive galaxy is present at a separation of $\sim$148 kpc and at the same redshift as suggested by the SN spectral features. The progenitor of PTF10ops is most likely a very old star, possibly in a low metallicity environment, which affects its explosion mechanism and observational characteristics.  PTF10ops does not easily fit into any of the current models of either subluminous or normal SN Ia progenitor channels.
\end{abstract}

\begin{keywords}
supernovae: general -- supernovae: individual (PTF10ops)
\end{keywords}

\section{Introduction} \label{intro}

Type Ia supernovae (SNe Ia) are thought to result from the thermonuclear explosions of CO white dwarfs (WD) in binary systems that accrete matter from their companion stars until they become unstable close to the Chandrasekhar mass limit. However, to date the progenitor of a SN Ia has not yet been detected to confirm the theory. The potential to standardise the lightcurves of SNe Ia using the correlation between the decline rate of their lightcurve from maximum light and the luminosity at maximum was first realised by \cite{phi93}, and since then major advances have been made in further standardising their lightcurves to produce a very successful method of determining the cosmological parameters of the universe \citep[e.g.][]{rie98, per99, rie07, kes09, sul11a}. 

In the course of these studies, SNe Ia that do not obey the standard relations and have unusual photometric or spectral properties have also been discovered. It is very difficult to explain the diversity of SN events seen using only one progenitor model, which suggests that there could be more than one mechanism for creating SNe Ia. One class of unusual SNe Ia are those of the 91bg-like family \citep[e.g.][]{fil92, lei93, tau08}, which are characterised by under-luminous peak magnitudes ($\sim$2 mag fainter than for a normal SN Ia), rapidly declining lightcurves, a lack of a secondary maximum in the near-infrared bands, cool spectra with significant \TIii\ absorption features and lower than normal ejecta velocities. Similarly to normal SNe Ia, a correlation between the lightcurve width and peak luminosity of subluminous SNe Ia exists but the values of the fit for this relation are different to that of normal SNe Ia \citep{gar04, gon10}. Furthermore, 91bg-like SNe explode preferentially in early type galaxies \citep{how01, gal05}. 

It is currently unclear if the progenitors and explosion mechanism of 91bg-like objects are fundamentally different to those of normal SNe Ia. Various progenitor system models have been investigated, to try to determine if they originate from a distinct progenitor channel or are formed in a manner similar to normal luminosity SN Ia, but with much lower \nick\ masses and hence cooler spectra and narrower lightcurves. One possibility is the merger of two WD, each with a mass of $\sim$0.9 \msun, which has been modelled and shown to produce SN explosions with some characteristics similar to those of subluminous SNe Ia \citep{pak10}.

Even more recently, a number of unusual transients have been discovered that do not fit cleanly into any of the existing classes of SNe Ia, and most interestingly appear to be found preferentially at either large distances from their host galaxies or possibly in undetected, very faint host galaxies.  SN 2005E \citep{per10} was a Ca-rich SN Ib that exploded at a distance of $\sim$22.9 kpc from its S0/a host galaxy and is suggested to originate from a low mass, old progenitor such as a He-accreting WD in a binary system. PTF09dav \citep{sul11} was a peculiar very faint SN Ia discovered at a distance of $\sim$41 kpc from its spiral host galaxy and no progenitor model has been found to explain its combined properties of low velocity spectral features, very low luminosity and strong lines of \Scii\ and \Mgi. PTF09dav will be discussed in more detail in Kasliwal et al (in prep.), along with two other faint SNe (PTF10iuv and PTF10bij) that were discovered far from their host galaxies. Another SN discovered at a large distance (33.7 kpc) from its S0/a host galaxy is the peculiar SN Ia 2006bt \citep{fol10}. It displayed the broad and slow lightcurve of a normal SN Ia and was marginally under-luminous, while its spectra were found to be most similar to those of subluminous 91bg-like SNe, displaying \TIii\ features and its $i$ band lightcurve lacked the secondary maximum that is seen in normal SNe Ia. Its position and its host galaxy type suggest that SN 2006bt came from a very old progenitor star, although the exact progenitor model to explain its properties is still not understood.

In this paper, we present the discovery and analysis of the SN Ia, PTF10ops, which was discovered by the Palomar Transient Factory \citep[PTF,][]{rau09, law09}. PTF10ops joins these new objects that are found at large distances from their host galaxies, and at a projected separation of $\sim$148 kpc from its host galaxy, PTF10ops claims the title of most remote SN discovered to date. PTF10ops also has the unusual combination of subluminous SN Ia like spectra but a lightcurve width typical of normal SNe Ia along with a low \nick\ mass.

 It is only with the advent of new wide area, optical surveys, such as the PTF, that scan the sky at a number of different cadences, that the discovery of these remote objects has been made possible. Two of the major advantages of this untargeted, scanning search strategy is that there is no selection bias toward massive host galaxies, and the high cadence repeat observations are optimal for the discovery of fast, subluminous events \cite[e.g.][]{kas10}. Therefore, due to the PTF and other modern sky surveys, the rate of discovery of SN transients that are located in faint host galaxies or far from their potential hosts is now increasing significantly and includes unusual SNe such PTF10ops which will be detailed here.

\section{Observations and Data Reduction}
PTF10ops was discovered by the PTF using the Galaxy Zoo Supernovae citizen science project\footnote{PTF10ops discoverers: Aleksandar,
Henryk Krawczyk, ciberjohn, Giovanni Iezzi, Elisabeth Baeten, Sarah Zahorchak, Graham Dungworth, John P Langridge, Marek, Robert Hubbard, Volatile}  \citep{smi11} on 2010 July 11.4 (UT dates are used throughout) at RA: 21:47:33.57, dec.: +05:51:30.3 (J2000) using the PTF search telescope, the Samuel Oschin 48-in telescope (P48) located at the Palomar Observatory. A spectrum was obtained on 2010 July 14.4 at the Palomar Hale 200in (P200) with the Double Beam Spectrograph \cite[DBSP;][]{oke82}, which showed PTF10ops to be a young SN Ia at a redshift of $z\sim0.06$. The spectral matching code \textsc{superfit} \citep{how05} was used to compare this spectrum to other SN spectra and was found to be most similar to the `normal' SN Ia, SN 1994D at a phase of $-8$ d \citep{ric95}. No narrow interstellar \Nai\ D lines were detected in the SN spectra, which suggests a low host galaxy extinction at the position of the SN  \cite[although see][]{poz11}, as would be expected for such a remote location. Therefore, no host galaxy extinction correction has been applied to the data. 

Due to its early discovery epoch, PTF10ops was selected as a suitable candidate for the non-disruptive Target of Opportunity program (GO 11721, P.I.: Ellis) on the Hubble Space Telescope (\textit{HST}) using the Space Telescope Imaging Spectrograph (STIS). The aim of this program was to obtain near-ultra violet (UV) spectra of SNe Ia near maximum light \citep{coo11}. PTF10ops was observed with the \textit{HST}+STIS on 2010 July 26.3 and surprisingly, given its earlier `normal' optical spectrum, was found to display the spectral signatures of subluminous SNe Ia such as strong lines of \TIii. Due to its \textit{HST} UV observation, a campaign of photometric and spectroscopic follow-up observations was scheduled and these observations are detailed below.

\subsection{Photometry}

Photometric observations of PTF10ops were obtained in \textit{gri} bands at the robotic 2-m Liverpool Telescope (LT) located at the Roque de Los Muchachos Observatory on La Palma. The field of PTF10ops was also monitored in the \textit{R} band by the P48.  The field of PTF10ops is in the Sloan Digital Sky Survey \cite[SDSS;][]{yor00} and a direct calibration of the photometry could be performed to the SDSS photometric system, close to the AB system \citep{oke83}. As will be detailed in Section \ref{host_galaxy}, no host galaxy was detected at the position of the SN down to an apparent magnitude limit of $r\geq25.1$ using the William Herschel Telescope (WHT) with the auxiliary-port camera (ACAM). Therefore, it was not necessary to wait for a reference image to subtract off any residual flux from the host galaxy when analysing the lightcurve.

Aperture photometry was used to measure a sequence of tertiary standards in the field of the SN to determine a zeropoint for each image. For the LT data, the colour terms to correct to the SDSS system are smaller than the errors on the photometry. The colour terms for the P48 are non-negligible ($\sim$0.2*($R-I$) in the $R$ band) but we have chosen to present the magnitudes in the natural P48 system and so a colour term has not been applied 
\cite[for more details see][]{sul11}. The SN magnitude was measured on each epoch using PSF photometry and calibrated to the SDSS system using the previously calculated zeropoints, along with an aperture correction to correct for the difference between the aperture in which the tertiary standards were measured and the fitted PSF. The optical magnitudes and their errors (combining statistical and calibration errors) for PTF10ops are detailed in Table \ref{opt_phot}. 

\begin{table*}
 \caption{The optical magnitudes and associated errors for PTF10ops are detailed, along with the telescopes used.}
 \label{opt_phot}
 \begin{tabular}{@{}lccccccccccccccccccccccccccccccc}
  \hline
  \hline
MJD$^a$&$g$& $r$ &$R$& $i$ & Telescope \\
\hline
  55382.5  &   &  &  22.184$\pm$0.490 &     &P48\\
  55388.4  &   &  &  20.233$\pm$0.081 &     &P48 \\
  55388.4  &   &  &  20.229$\pm$0.073 &     &P48\\
    55392.0   &    19.837$\pm$0.039   &   19.630$\pm$0.046   &   &  19.713$\pm$0.046  &      LT     \\
  55394.0    &     19.557$\pm$0.038  &   19.380$\pm$0.035   & &   19.656$\pm$0.056  &   LT  \\
  55396.2    &   19.376$\pm$0.080  &   19.267$\pm$0.067   & &   19.587$\pm$0.126   &  LT  \\
  55397.5  &   &  &  19.116$\pm$0.050 &     &P48\\
    55398.0    &    19.338$\pm$0.074 &   19.046$\pm$0.046   & &   19.250$\pm$0.070  &   LT  \\
  55398.4  &   &  &  19.083$\pm$0.033 &     &P48 \\
  55398.4  &   &  &  19.102$\pm$0.035 &     &P48 \\
     55402.0    &  19.463$\pm$0.233    &   18.955$\pm$0.126    &&    19.008$\pm$0.118   &  LT  \\
  55410.0    &   19.869$\pm$0.067  &   19.157$\pm$0.041   & &   19.415$\pm$0.057  &   LT \\
  55410.5  &   &  &  19.231$\pm$0.050 &     &P48  \\
  55410.5  &   &  &  19.330$\pm$0.073 &     &P48  \\
   55411.1    &   19.965$\pm$0.068   &   19.191$\pm$0.043   & &   19.350$\pm$0.047  &   LT   \\
  55414.1    &     20.408$\pm$0.076 &   19.394$\pm$0.052   & &   19.433$\pm$0.049  &   LT  \\
  55417.1    &   20.650$\pm$0.142   &   19.388$\pm$0.075   & &   19.477$\pm$0.070  &   LT\\
  55417.4  &   &  &  19.544$\pm$0.079 &     &P48  \\
   55423.0    &    &   19.650$\pm$0.097   & &   19.671$\pm$0.065 &   LT \\
  55423.3  &   &  &  19.751$\pm$0.057 &    &P48  \\
  55423.3  &   &  &  19.797$\pm$0.057 &     &P48  \\
  55426.3  &   &  &  19.973$\pm$0.094 &     &P48  \\
  55426.4  &   &  &  19.905$\pm$0.061 &     &P48  \\
  55430.2  &   &  &  20.179$\pm$0.190 &     &P48  \\
  55468.0   &  &   21.136$\pm$0.061 & & 21.186$\pm$0.096&LT\\
55685.2&&$\geq$25.1&&&WHT\\
\hline
 \end{tabular}
 \begin{flushleft}
 $^a$Modified Julian Date, JD--2400000.5. \\
\end{flushleft}
\end{table*}

\begin{figure}
\includegraphics[width=8cm]{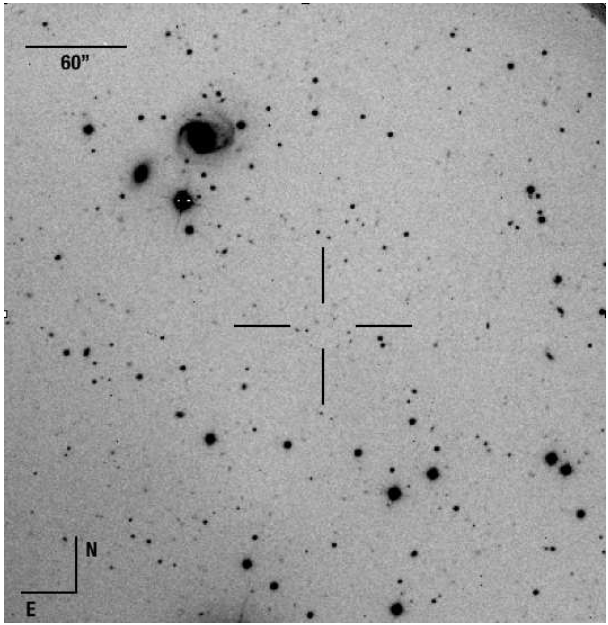}
\caption{An image of the field of PTF10ops (located at the centre of the crosshairs) taken with the WHT+ACAM with 0.25 arcsec per pixel. The largest galaxy located in the upper left quadrant has a projected separation from PTF10ops of 115.2$''$ (1.92$'$), which at $z_{cmb}$ = 0.062 corresponds to 148.3 kpc. }
\label{galaxy}
\end{figure}

\subsection{Spectroscopy}

Optical spectra of PTF10ops (including the classification spectrum) were obtained at 7 epochs ranging from $-8$ d to +102 d with respect to \textit{B} band maximum and are detailed in Table \ref{opt_spec}. The classification spectrum taken on 2010 July 14.4, along with a spectrum on 2010 August 13.4, were obtained at the P200+DBSP 
using the 600l/4000~\AA\ (blue) and 158l/7500~\AA\ (red) gratings. A spectrum was obtained with the Supernova Integral Field Spectrograph \cite[SNIFS;][]{lan04} on the University of Hawaii 2.2-m (UH), operated by the Nearby Supernova Factory on 2010 July 21.4. Two spectra were obtained with the WHT and Intermediate dispersion Spectrograph and Imaging System (ISIS) on 2010 August 02.1 and 2010 August 03.1 using the R158R (red) and R300B (blue) gratings.
A spectrum was also obtained at the Lick Observatory Shane 3-m telescope using the KAST double spectrograph \citep{mil93} on 2010 August 11.3 with the 600/4310 (blue) and 300/7500 (red) gratings. Finally, a near-nebular phase spectrum of PTF10ops was obtained with the Keck-1 using the Low-Resolution Imaging Spectrometer \cite[LRIS;][]{oke95} on 2010 November 01.0 using the 400l/3400\AA\ grism.

The optical spectra were reduced using a custom pipeline based on standard spectral reduction procedures in \textsc{iraf} and \textsc{idl}.The 2D spectra were de-biased and flat-fielded before extraction. The extracted spectra were wavelength calibrated using arc-lamp exposures and instrumental response functions were obtained from observations of spectrophotometric standards to perform the flux calibration. The flux calibration was subsequently checked using coeval photometry if available and adjusted to match these magnitudes.

A near-UV spectrum was obtained with \textit{HST}+STIS (GO 11721, P.I.: Ellis) on 2010 July 26.3, listed in Table \ref{opt_spec}. This spectrum was downloaded from the Space Telescope Science Institute (STScI) using the on-the-fly reprocessing pipeline (OTFR) to give a fully calibrated 1D spectrum. 

\begin{table*}
 \caption{Log of the optical and UV spectral observations of PTF10ops.}
 \label{opt_spec}
 \begin{tabular}{@{}lccccccccccccccccccccccccccccccc}
  \hline
  \hline
MJD&Phase (d)$^b$& Telescope+Instrument& Range (\AA)\\  \hline

55391.4&-7.3    &P200+DBSP  & 3500--9300\\
55398.5&-0.6    & UH+SNIFS&  3300--9700 \\
55403.3&+4.0  &HST+STIS&2800--5700\\
55410.1&+10.4 & WHT+ISIS& 3000--11200 \\
55411.1&+11.3 &WHT+ISIS &  3000--11200 \\
55419.3&+19.0 & LICK+KAST&3400--9900&   \\
55421.3&+20.9 &P200+DBSP & 3500--9300 \\
55501.3&+96.2 &Keck 1+LRIS &3500--7500  \\
\hline
 \end{tabular}
 \begin{flushleft}
  $^a$Days in rest-frame relative to $B$ maximum, MJD 55399.1. \\
\end{flushleft}
\end{table*}

\subsection{Host Galaxy}
\label{host_galaxy}

Figure \ref{galaxy} shows a late time $r$ band image of PTF10ops, which was obtained at the WHT+ACAM and no galaxy or the SN was detected down to a limit of $r\geq25.1$ mag for a 3$\sigma$ detection. At the redshift suggested by the SN spectrum of z$\sim$0.06, this corresponds to an absolute magnitude limit of $-12.0$ mag. \cite{arc10} studied a sample of 72 core-collapse SNe found with the PTF and found the faintest host galaxy to be that of PTF10dk with an $r$ band magnitude of $-14.28$ mag. However, despite the relatively deep limit, a faint dwarf host galaxy cannot be completely ruled out. 

Another possibility for the host galaxy of PTF10ops is the massive spiral galaxy seen in the upper left quadrant of Fig \ref{galaxy}. PTF10ops is located at a distance of 115.2$''$ from this spiral galaxy, SDSS J214737.86+055309.3 (RA: 21:47:37.86, dec.: +05:53:09.3 (J2000)). A spectrum of this galaxy was obtained at the WHT+ISIS with the R300B and R158R gratings on 2010 October 30.8 and its features such as Ca H\&K and Na ID are consistent with \textit{$z_{hel}$} = 0.061$\pm$0.001. This redshift is consistent with the estimated SN redshift and corresponds to \textit{$z_{cmb}$} = 0.062 (distance modulus of 37.12$\pm$0.04, H$_{0}$ = 70 km s$^{-1}$ Mpc$^{-1}$ is used throughout). At this redshift, this implies a physical separation between PTF10ops and the centre of this galaxy of $\sim$148 kpc. The apparent \textit{ugriz} magnitudes from the SDSS DR8 of this galaxy are 18.03, 16.13, 15.22, 14.77 and 14.42 mag respectively. SED fitting of the host photometry to a series of galaxy templates was performed using the \textsc{z-peg} photometric redshift code of \cite{leb02} and the method of \cite{sul10} to estimate the galaxy stellar mass. It is estimated to be log(M/\msol) = 11.18$^{+0.04}_{-0.25}$, which places it at the upper end of stellar masses of the host galaxies of SNe Ia \citep{sul10}.

\section{Photometric Analysis}
\subsection{Lightcurve analysis}

\begin{figure*}
\subfigure{\includegraphics[width=8.7cm]{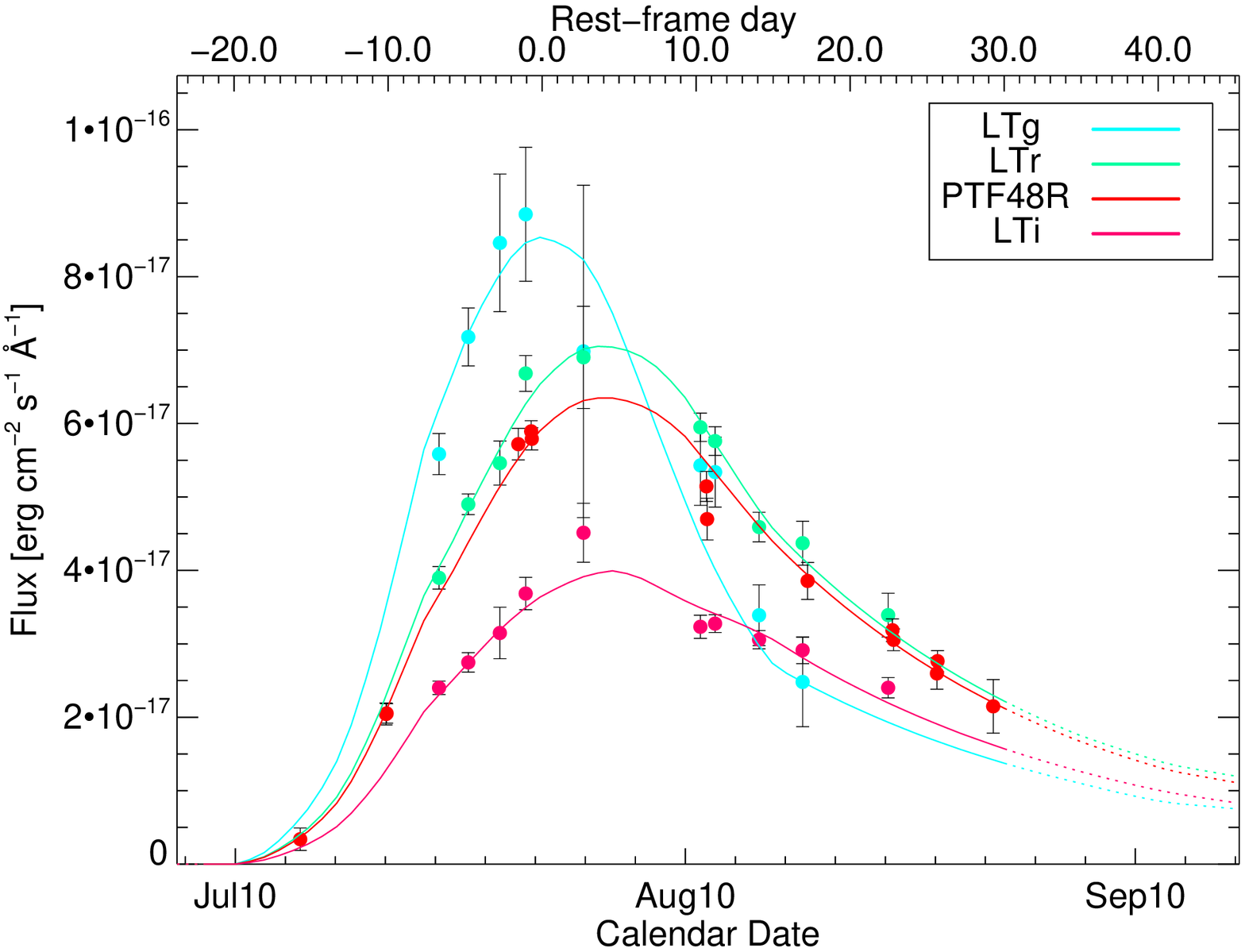}}
\subfigure{\includegraphics[width=8.7cm]{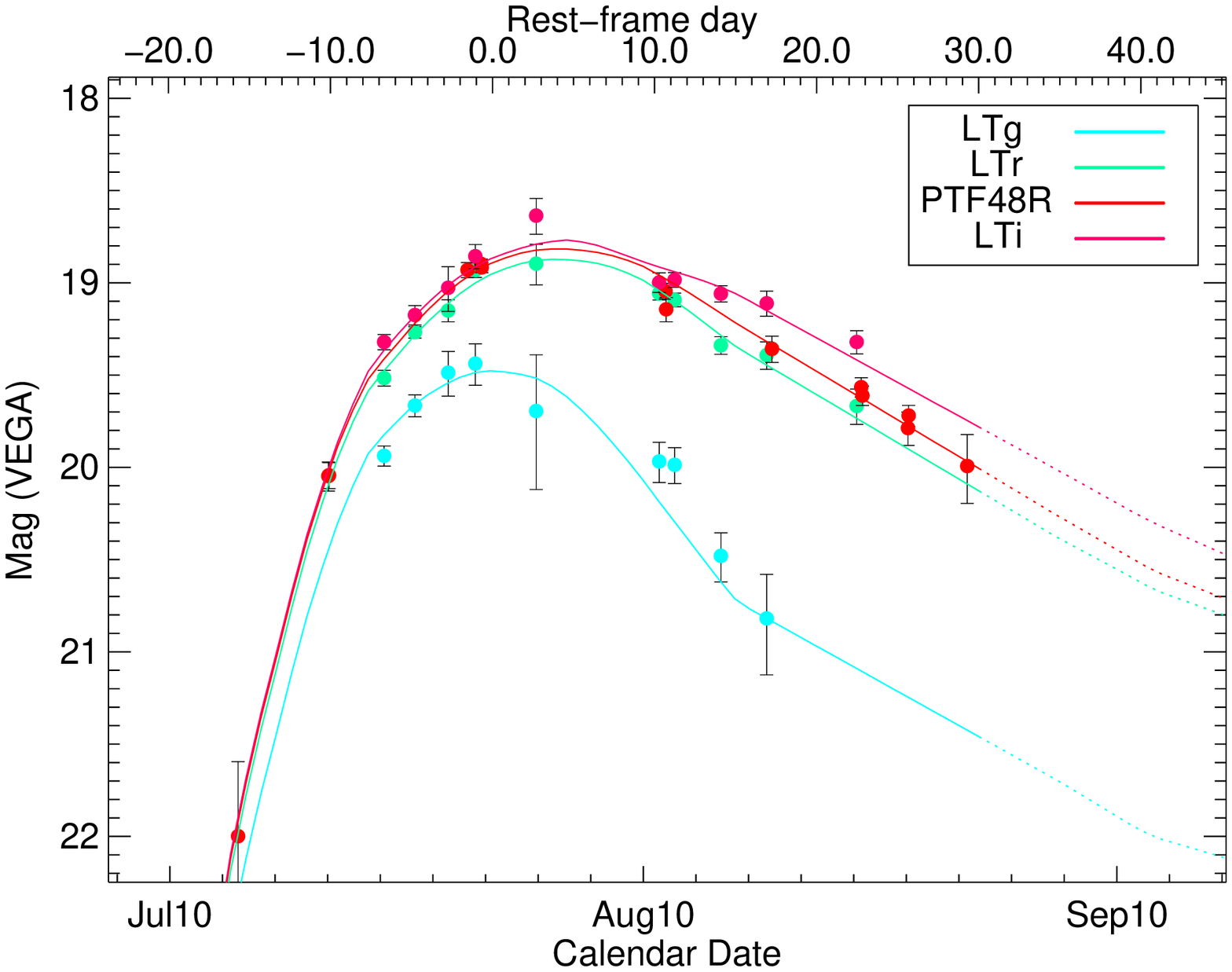}}
\caption{The observed lightcurve of PTF10ops using data taken with the LT and the P48 in flux space (left) and magnitude space (right). The solid lines are the best-fitting SiFTO lightcurve for data up to +30 d using a subluminous SN Ia spectral template. The reduced \protect $\chi^{2}$ of the fit is 1.35 for 37 dof.}
\label{10ops_fit91bg}
\end{figure*}

The stretch and peak luminosity of a SN Ia can be determined by applying a lightcurve fitter to its data. In the case of PTF10ops, we have used the SiFTO lightcurve fitter \citep{con08}. The inputs to SiFTO for PTF10ops are the observed photometry detailed in Table \ref{opt_phot}, along with its redshift and extinction. SiFTO uses a time series of spectral templates that are adjusted to recreate the observed colours of the SN photometry at each epoch, while also adjusting for Galactic extinction and redshift. The time-axis of the lightcurve synthesised from the template spectra (i.e width of the lightcurve) is adjusted to match that of the SN, using a `stretch' value relative to that of the template. This fit can then be used to interpolate the peak magnitude in a chosen rest-frame filter. The outputs of the SiFTO fit include the stretch, date of maximum magnitude, goodness of fit and absolute magnitude in a chosen filter.

PTF10ops displays characteristics that are found in both normal luminosity (normal stretch) and subluminous (low stretch) SN Ia events so we have investigated both normal and subluminous spectral templates to parameterise the lightcurve and determine which is a better fit. The subluminous template\footnote{http://www.lbl.gov/$\sim$nugent/nugent$\_$templates.html} \citep{nug02} is based on two subluminous SNe Ia, SNe 1991bg \citep{fil92, lei93} and 1999by \citep{gar04}, while the normal template is based on the sample of SNe Ia of \cite{hsi07} with stretches in the range 0.6--1.2. It should be noted that the stretch values quoted here for both templates are with respect to the average value of the respective templates and are not directly comparable.

The results of the SiFTO fitting routine for PTF10ops for a subluminous template are a relative stretch of 1.51$\pm$0.02 with a reduced ${\chi}^2$ of 1.35 for 37 degrees of freedom (dof). The normal template gives a stretch of 1.09$\pm$0.02 with a reduced ${\chi}^2$ of 4.08 for 33 dof.  The ${\chi}^2$ value of the normal fit is significantly worse than for the subluminous fit. This worse fit is mainly due to the poor agreement between the normal template and $g$ band data, where the evolution of the lightcurves is significantly different from the normal template. This is most likely caused by the strong \TIii\ features that are present in subluminous SN spectra at wavelengths covered by the $g$ band filter. We have investigated the small discrepancy that is seen in the $g$ band between the data and the model at $\sim$10 d and the most plausible explanation is that the photosphere of PTF10ops is hotter at this time than that of the subluminous template. The subluminous template displays a much better agreement overall and therefore, the subluminous template has been chosen as the most suitable model and is used in the following analysis. 

Figure \ref{10ops_fit91bg} shows the observed lightcurve of PTF10ops in both flux and magnitude space, along with the best-fitting SiFTO lightcurves using the subluminous SN Ia spectral template. The data are well fitted by this template, including an early 2.5$\sigma$ $R$ band detection from $-15.6$ d (rest-frame). This means that the rise-time from explosion to maximum light can be well constrained. 

The subluminous template used in the fit has a rise-time, $t_{r}$ to $B$ band maximum of 13.0 d, which means for the stretch of PTF10ops of 1.51$\pm$0.02, the rise-time is found to be 19.6$\pm$0.3 d (statistical error only). The rise-time was also calculated using the early time $R$ band data from $-16$ to $-10$ d, by applying the method of \cite{rie99} and $t_{r}$ was found to be 20.0$\pm$2.2 d. This value is consistent with the value obtained from the lightcurve fit but has a larger associated error, so the rise-time used in the rest of the paper is 19.6$\pm$0.3 d. This value is toward the higher end of the range of rise-times to $B$ band peak brightness seen for normal SNe Ia, of 13--23 d before correction for stretch \citep{hay10}. Gonz\'alez-Gait\'an (in prep.) also find a similar rise-time range for normal SNe Ia of 14-21 d, while for subluminous, low stretch SNe they find a rise-time range of 10--14 d. Therefore, the rise-time of PTF10ops is significantly longer than those of subluminous SNe Ia and instead lies in the normal SNe Ia range.

The calendar date of $B$ band max, obtained from the SiFTO fit, is 2010-07-22.1, which corresponds to an MJD of 55399.1. The peak rest-frame apparent \textit{B} and $V$ magnitudes of PTF10ops are calculated to be 19.35$\pm$0.04 and 19.01$\pm$0.02 respectively. The magnitudes reported in this paper are in the Vega system using the filter responses from \cite{bes90} and are corrected for Galactic extinction, assuming an E(\bv) of 0.064 \citep{sch98} and the \cite{car89} extinction law (R$_V$=3.1 is used throughout). Using a distance modulus of 37.12$\pm$0.04, the absolute $B$ and $V$ magnitudes at maximum of PTF10ops are calculated to be $-17.77\pm$0.04$\pm$0.04 and $-18.11\pm$0.02$\pm$0.04, respectively, where the errors are statistical and systematic respectively. The absolute magnitude of PTF10ops is $\sim$1.5 magnitudes fainter than the average $B$ band luminosity of normal SNe Ia of $-$19.3 mag \citep{ben05}, which suggests that based on its luminosity, PTF10ops is most similar to the 91bg-like subclass of SNe Ia.

The (\bv)$_{{max}}$ value of PTF10ops at maximum $B$ band magnitude is 0.34$\pm$0.04, after correction for Galactic extinction. Subluminous 91bg-like objects generally have redder optical colours than normal SNe Ia at maximum $B$ band light, with values typically in the range, 0.4--0.7 mag, compared to $\sim$0 mag for normal luminosity SNe Ia \citep{tau08}. The (\bv)$_{\textrm{max}}$ value of PTF10ops is intermediate between normal and subluminous SN values, which can be explained by the \TIii\ absorption features that are present but that are not as strong as for 91bg-like objects.

The value of the \cite{phi93} decline rate parameter, $\Delta$M$_{15}$(\textit{B}) is estimated by determining the change in $B$ band magnitude between maximum and 15 d after maximum in the rest-frame of the SN. For PTF10ops, this results in a value of $\Delta$M$_{15}$(\textit{B}) of 1.12$\pm$0.06, which is a typical value for a normal SN Ia. Using the Phillips relation, this corresponds to a peak absolute $B$ band magnitude of $-19.47$, which is 1.7 magnitudes brighter than actually observed for PTF10ops. \cite{tau08} also determined a (different) relation between the peak absolute $B$ band magnitude of a subluminous SN and its $\Delta$M$_{15}$(\textit{B}) value. However, this relation only holds for `standard' subluminous SNe Ia with $\Delta$M$_{15}(\textit{B})\geq1.69$ and cannot be applied to PTF10ops.  This clearly shows that PTF10ops is an outlier in both normal and subluminous SNe Ia comparisons of $\Delta$M$_{15}$(\textit{B}) and absolute $B$ band magnitude. \cite{kas08} found a correlation for subluminous SNe Ia between the peak $B$ band magnitude and the time of intercept of the early and late time $B$ lightcurve slopes, $t_b$, as first defined in \cite{psk84}. Using the relation of \cite{kas08} and a value of $t_b$ of $\sim$16 d measured from the lightcurve of PTF10ops, we estimate a peak $B$ band magnitude of $\sim$17$\pm$1 mag, which is in much better agreement with the measured value of PTF10ops compared to those that rely on the more commonly used parameter, $\Delta$M$_{15}$(\textit{B}).

\begin{table*}
 \caption{Comparison of the properties of a sample of SNe Ia.}
 \label{table_comp1}
 \begin{tabular}{@{}lccccccccccccccccccccccccccccccc}
  \hline
  \hline
  SN & 	M$_{\textit{B}}$$^a$(mag)  & (\bv)$_{\textrm{max}}$& Stretch$^b$ & $\Delta$m$_{15}$(\textit{B}) & $\mu$ &References$^c$\\
  \hline
SN 2006bt&-18.94$\pm$0.06&0.12$\pm$0.04&1.57$\pm$0.02&1.09$\pm$0.06&35.69&\cite{fol10} \\
SN 1986G & -18.30$\pm$0.13&0.31$\pm$0.01&1.04$\pm$0.01&1.66$\pm$0.06&27.67&\cite{tau08, phi87} \\
SN 2007au&-18.20$\pm$0.03&0.20$\pm$0.04&1.16$\pm$0.04&1.95$\pm$0.11&34.69&\cite{sul11, hic09a}\\
PTF10ops &-17.77$\pm$0.06&0.34$\pm$0.04&1.51$\pm$0.02&1.12$\pm$0.06 &37.12&this work \\
SN 2005bl&-16.72$\pm$0.09&0.60$\pm$0.02&0.96$\pm$0.01&1.67$\pm$0.06&35.16&\cite{sul11, tau08} \\
SN 1991bg & -16.71$\pm$0.03&0.66$\pm$0.02&0.83$\pm$0.04&1.94$\pm$0.10&30.90&\cite{sul11, tau08}\\
SN 2005ke&-16.54$\pm$0.20&0.66$\pm$0.07&1.08$\pm$0.01&1.66$\pm$0.14&31.42&\cite{sul11, hic09a} \\
\hline
 \end{tabular}\\
 \begin{flushleft}
 $^a$Corrected for Galactic extinction but not host galaxy extinction, except in the case of SN 1986G, which had substantial host galaxy extinction of E(\bv)=0.67. \\
 $^b$The stretch values are relative to our subluminous SN Ia template.\\
 $^c$These are the references for the observed parameters of the SN sample.\\
\end{flushleft}
\end{table*}

To investigate further the unusual photometric behaviour of PTF10ops, its properties were compared to those of a sample of SNe, detailed in Table \ref{table_comp1}. This sample of SNe includes subluminous SNe Ia with narrow, fast lightcurves, as well as the peculiar SN 2006bt. The literature was searched for more events with similar properties to PTF10ops but none with its combination of sub-luminosity and slow lightcurve decay were found. The magnitudes for the SNe in Table \ref{table_comp1} have been corrected for Milky Way extinction but no host galaxy extinction as been applied, except in the case of SN 1986G for which there was a substantial host galaxy extinction of E(\bv) of 0.67 \citep{tau08}. The stretches of these SNe are given with respect to the subluminous SiFTO template, described above. 

\begin{figure}
\includegraphics[width=8cm]{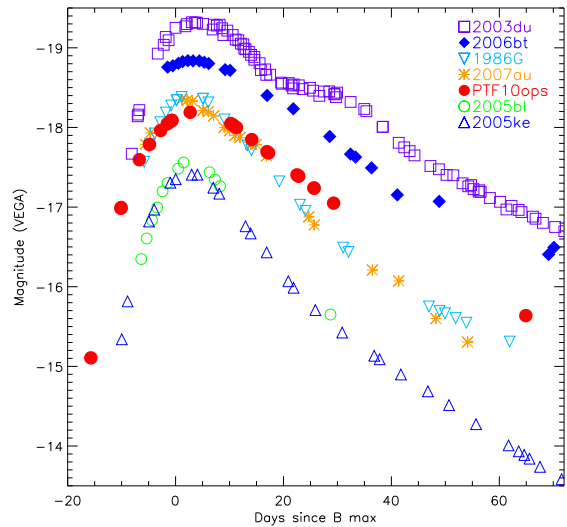}
\caption{A comparison of the absolute \textit{r} band lightcurve of PTF10ops to a sample of other SNe Ia: SNe 2003du, 2006bt, 2006ax, 2005ke, 2005bl and 1986G. The lightcurves have been corrected for Milky Way extinction using \protect \cite{sch98} and are in the SN rest-frames. The lightcurve of SN 1986G has also been corrected for its significant host galaxy extinction of E(\bv)=0.67 (R$_V$=3.1). The P48 $R$ band data of PTF 10ops and the $R$ band lightcurve of SN 1986G have both been transformed to the LT $r$ band for comparison purposes. 
}
\label{r_comp_lc}
\end{figure}

Figure \ref{r_comp_lc} shows a comparison of the rest-frame $r$ band lightcurve of PTF10ops with the sample of the SNe detailed in Table \ref{table_comp1}, along with the normal SN Ia 2003du \citep{sta07}, which displayed a clear shoulder in its \textit{r} band lightcurve. The $r$ band lightcurve of PTF10ops is subluminous compared to SN 2003du and does not display a prominent shoulder in its \textit{r} band lightcurve, similar to the lightcurves of subluminous SNe Ia (see Section \ref{bolo_nick} for more on this). It is also worth highlighting that the subluminous SNe 1986G and 2007au both have similar luminosities to PTF10ops, but neither display the large relative stretch value of PTF10ops. 

Based on its lightcurve shape, lower than normal luminosity and slow lightcurve decline rate, PTF10ops is most similar to SN 2006bt.  To further investigate their similarities, $B$ and $r$ band data of SN 2006bt were analysed using the SiFTO lightcurve fitter for both the subluminous and normal templates in a similar manner to PTF10ops. The best fit is found to be the normal template, although the $r$ band appears more similar to the subluminous $r$ band template. The stretches of SN 2006bt for the subluminous and normal templates are 1.57$\pm$0.02 with a reduced ${\chi}^2$ of 6.83 for 27 dof and 1.03$\pm$0.04 with a reduced ${\chi}^2$ of 2.14 for 21 dof respectively. Therefore, despite their similar lightcurve shapes (lack of a secondary peak yet large stretch values) PTF10ops and SN 2006bt are best fit with subluminous and normal lightcurve templates respectively. This reflects the different strengths of spectral features, when compared to those present in the template spectra. Since the subluminous spectral template is based on only two SNe, the different best fits for PTF10ops and SN 2006bt really describe how similar (or not) to these particular objects they are. The peak $B$ band magnitude of SN 2006bt is 1.17 brighter than for PTF10ops, which can not be solely explained by the difference in optical colour that is seen at peak magnitude.  In the future it would be interesting to expand the current low stretch template to include more subluminous spectra with a larger range in subluminous feature strengths. 

\subsection{Bolometric luminosity and \nick\ mass}
\label{bolo_nick}

The bolometric luminosity of PTF10ops can be estimated using the SiFTO lightcurve fits. To calculate the bolometric luminosity of a SN, a bolometric correction must be applied to convert from the bands observed to the full $UVOIR$ lightcurve. We calculated the bolometric luminosity of PTF10ops using a method broadly similar to \cite{how09} by using a subluminous template spectrum at the appropriate epoch (flux normalised and colour corrected to match the SiFTO lightcurve fit) and extrapolated to the blue via a comparable model from \cite{nug95} and into the infrared via a 9000 K blackbody spectrum. This spectrum was then integrated to obtained the bolometric luminosity at each epoch. The observed spectra of PTF10ops could be used instead of the template spectrum but due to the good fit of the data with the template and the template's wider wavelength coverage, the template spectrum was chosen. The peak bolometric luminosity, $L_{bol}$ of PTF10ops is found in this way to be 3.8$\pm$0.1 $\times$ $10^{42}$ erg s$^{-1}$, which occurs at $\sim$1 d post $B$ band peak.

Assuming the lightcurve is powered by the radioactive decay of \nick, the mass of \nick\ of a SN Ia can be estimated using the rise-time and bolometric luminosity at $B$ band maximum in the following equation \citep{arn82},
\begin{equation}
M_{Ni} = \frac {L_{bol}}{\alpha \dot{S}(\tau_r)}
\end{equation}
where $\dot{S}$ is the radioactive luminosity per solar mass of \nick\ from the decay of \nick\ to \cob\ at maximum light and $\alpha$ is the ratio of the bolometric to radioactive luminosities, which has a value of 1.2, given full $\gamma$-ray trapping at maximum \citep{nug95b}. $\dot{S}$ is defined as in \cite{nug95b} and is a function of $t_{r}$, where $t_{r}$ =  19.6$\pm$0.3 d. This results in a \nick\ mass estimate for PTF10ops of 0.17$\pm$0.01 \msun. This value is within the range of \nick\ masses (0.05--0.35 \msun) found for a sample of 30 subluminous SNe Ia (Gonz\'alez-Gait\'an, in prep.). 

\section{Spectroscopic Analysis}
\begin{figure*}
\includegraphics[width=10cm]{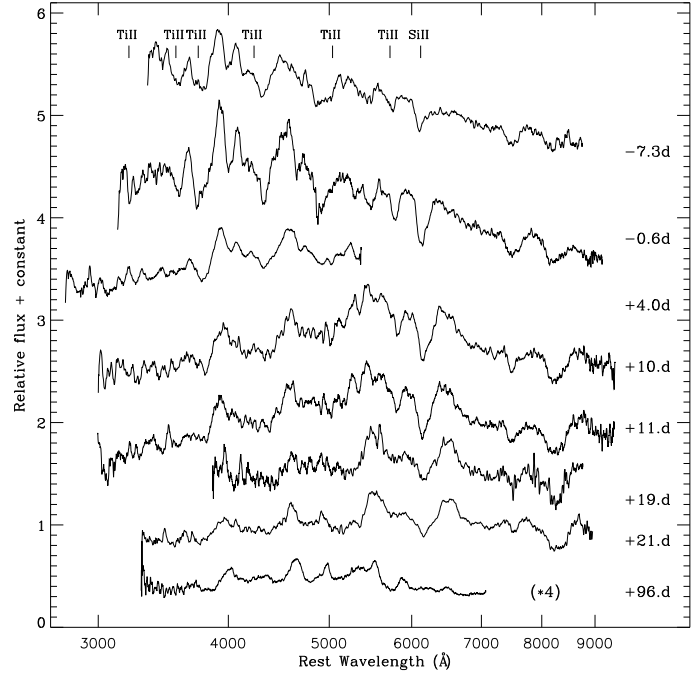}
\caption{Time evolution of the optical spectra of PTF10ops. The phases shown are relative to $B$ band maximum (rest-frame) and the spectra are in their rest-frame corrected using z$_{hel}=0.061$. The final spectrum has been multiplied by four for plotting purposes.}
\label{10ops_evo_spec}
\end{figure*}

\cite{kas06} modelled the secondary maxima in the NIR lightcurves of SNe Ia and found correlations between the strength and timing of the secondary maxima with the \nick\ mass, the degree of mixing, the mass of electron capture elements, the progenitor metallicity and the mass of intermediate mass elements. The luminosity of the secondary peak is expected to be small, as seen in the $i$ band lightcurve of PTF10ops, for SNe with low Fe masses and/or high degrees of mixing. The $i$ band lightcurve of PTF10ops was compared to the models of \cite{kas06} with different \nick\ masses (0.1--0.9 \msun\ in steps of 0.1 \msun) and found to most similar to the model with a \nick\ mass of 0.1 \msun, which is similar to the value obtained using the method of \cite{arn82}. This suggests that the ejecta are cool, which is also seen to be the case in the analysis of the spectra of PTF10ops. Similarly, varying the degree of outward mixing of \nick\ produces less prominent secondary maxima in the $i$ band for more homogenised compositions structures compared to models with lower values of mixing. 

The spectral evolution of PTF10ops is shown in Fig. \ref{10ops_evo_spec} for epochs in the range $-7$ d before to 96 d post $B$ band maximum light (rest-frame). The first spectrum of PTF10ops looks like that of a normal SN Ia with a prominent \SiII\ 6355~\AA\ absorption. However, the subsequent three spectra also display strong \TIii\ features (absorption near 3600, 4150, 4650 and 5000 \AA), which are seen in 91bg-like objects and are characteristic of a cool SN photosphere \citep{fil92}. The UV spectrum of PTF10ops obtained at +4 d also shows strong \TIii\ absorption features, which are not seen in the UV spectra of `normal' SNe Ia.

\begin{figure}
\includegraphics[width=8.5cm]{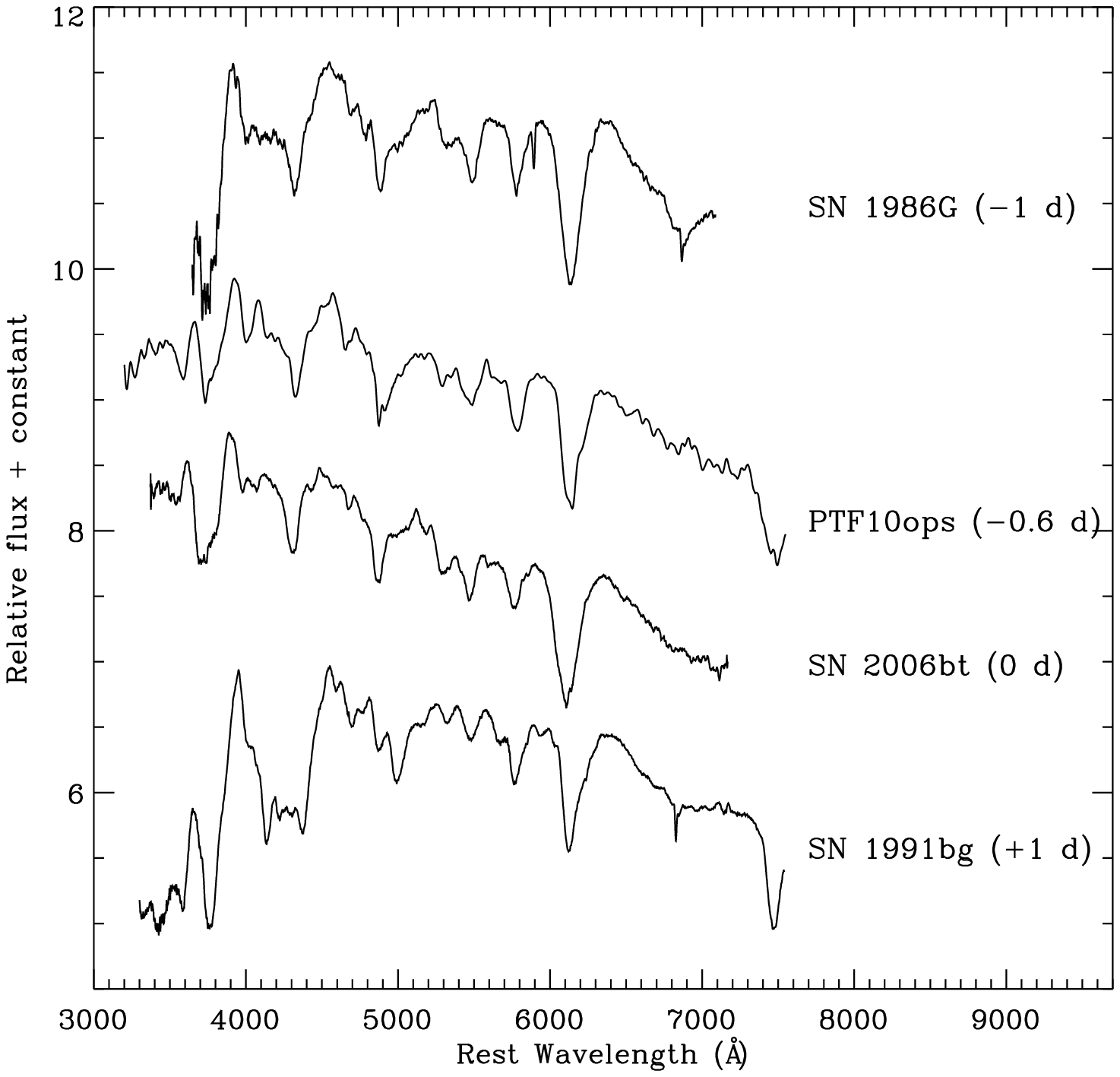}
\caption{Comparison of a $-0.6$ d spectrum of PTF10ops with SN 1986G, a peculiar subluminous SN Ia, SN 2006bt, a unusual SN Ia with similar properties to that of PTF10ops and SN 1991bg, the prototype of the subluminous SN Ia class. The spectra have been adjusted to their host velocities and corrected for MW extinction. (In the case of SN 1986G, it has also been corrected for host galaxy extinction of E(\bv)=0.67.) The phases shown are rest-frame, relative to $B$ band maximum.}
\label{10ops_comp_spec}
\end{figure}

\begin{figure}
\includegraphics[width=8.5cm]{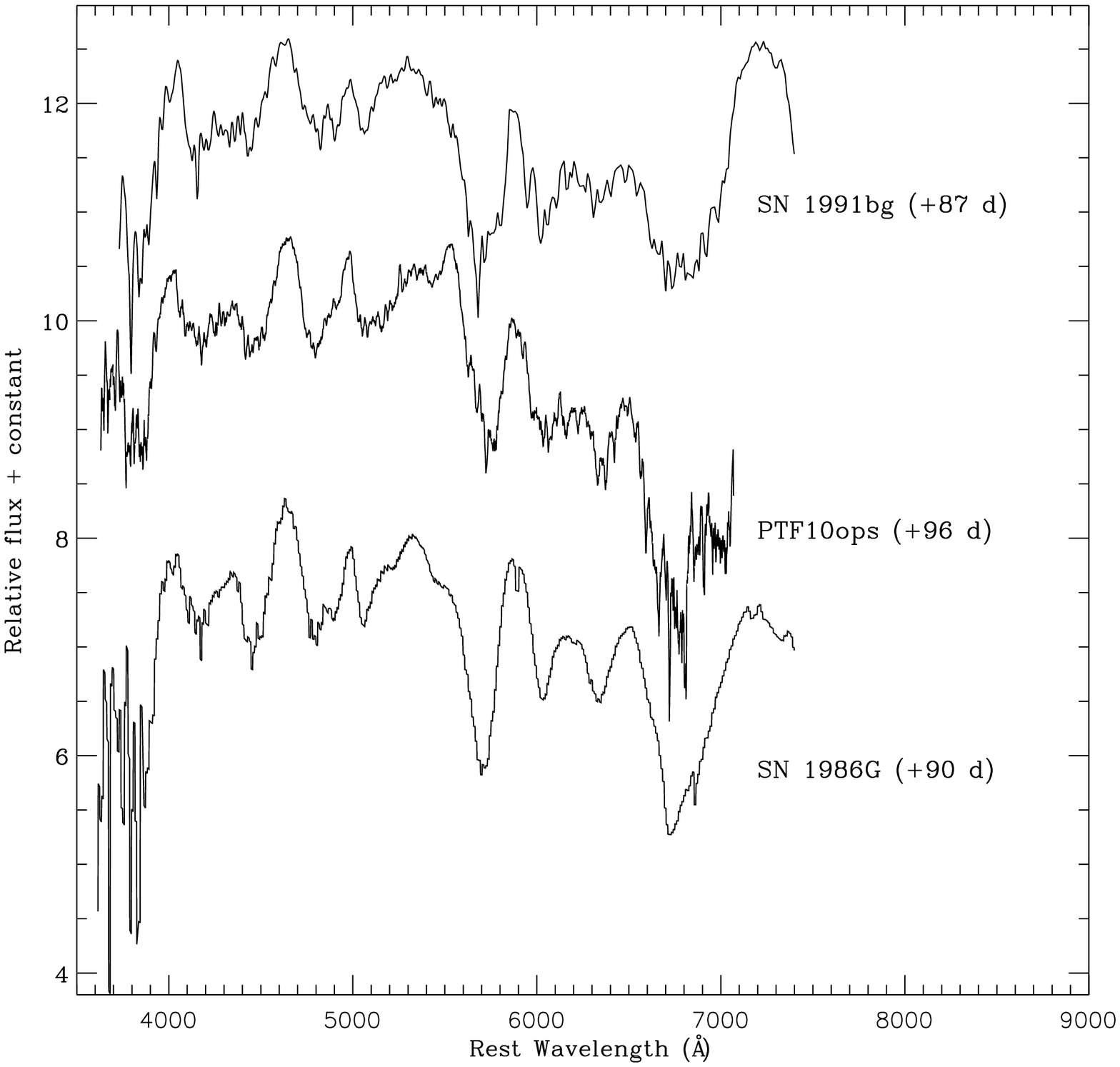}
\caption{Comparison of a +96 d spectrum of PTF10ops with SNe 1991bg and 1986G. The spectra have been adjusted to their host velocities and corrected for MW extinction. The phases shown are rest-frame, relative to $B$ band maximum.}
\label{10ops_comp_spec_late}
\end{figure}

In Fig. \ref{10ops_comp_spec}, the $-0.6$ d spectrum of PTF10ops is compared to those of SN 1986G (-1 d), SN 2006bt (0 d) and SN 1999bg (+1 d). SN 1986G was a subluminous SN Ia and a member of the 91bg-like class, although it displayed less prominent  \TIii\ lines, highlighting its intermediary phase between normal and subluminous events, while SN 2006bt was an unusual SN Ia like PTF10ops, with cool spectra but a broad lightcurve. The $-0.6$ d spectrum of PTF 10ops is most similar to that of SN 2006bt, with similar spectral features of approximately the same absorption depths, which is unsurprising given their combination of subluminous and normal SN Ia properties. The SN 1986G spectrum is also quite similar to both these spectra, while SN 1991bg has different features present along with much stronger \TIii\ absorption at $\sim$4150~\AA\ than seen in the other SNe. In Fig. \ref{10ops_comp_spec_late}, the late time, nearly nebular phase spectrum of PTF10ops is compared to those of SN 1991bg at +87 d and SN 1986G at +90 d. In this comparison, the spectral properties of PTF10ops appear intermediate to those of SNe 1991bg and 1986G.

\subsection{Spectral fitting}

For the photospheric phase spectra, the SN spectral fitting code \textsc{synapps} was used \citep{tho11}. \textsc{synapps} is based upon the well-known \textsc{synow} code \citep{fis00} and has the same assumptions and limitations. \textsc{synapps} automatically optimises over the input parameters, unlike the interactive \textsc{synow}, which required the user to adjust the parameters to obtain a best fit. The input parameters are an overall max. and min. photospheric velocity and blackbody photospheric temperatures, along with a line opacity profile, velocity limit for the opacity profile and a Boltzmann excitation temperature for each of the ions used in the fit.

\begin{figure*}
\includegraphics[width=10cm]{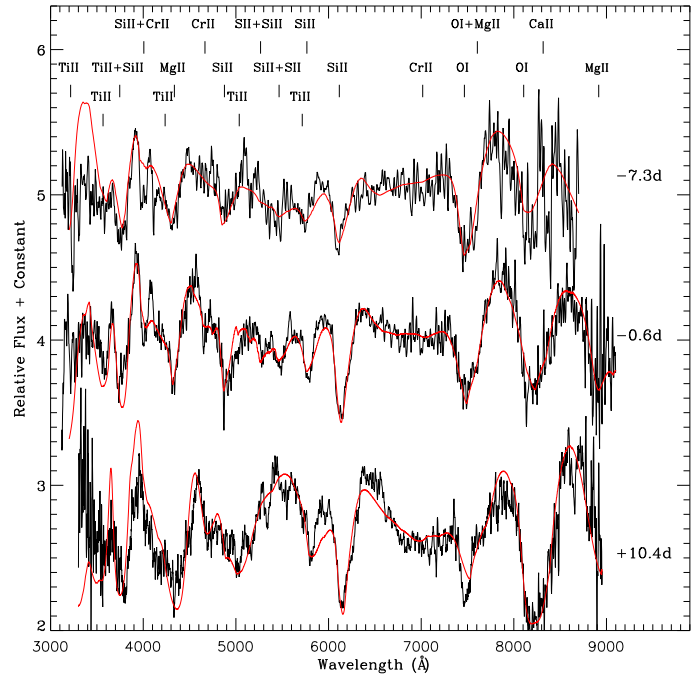}
\caption{The first three photospheric phase spectra of PTF10ops (black) at $-7.3$ d, $-0.6$ d and +10.4 d, along with their \textsc{synapps} fits (red). The spectra have been normalised by the continuum flux at each wavelength and the ions contributing to the major features are marked.}
\label{model_spec}
\end{figure*}

\textsc{synapps} was run on the first three photospheric phase spectra of PTF10ops as listed in Table \ref{opt_spec} and the resulting fits are shown in Figure \ref{model_spec}. Excellent agreement is seen between the models and observed spectra, with the models reproducing all the major spectral features. The ions used to produce the fits on the three epochs are made up of \Oinir, \Mgii, \SiII, \Suii, \Caii, \TIii, \Crii\ and \Feii. However, not all the lines were needed on each epoch. \Crii\ was found only to become important for the second and third epochs; it was not needed to produce the fits for the first spectrum at $-7.3$ d. Contributions from \Feii\ were only seen in the third spectrum at +10.4 d, while \Mgii\ and \Suii\ lines were found not to make a major contribution at this epoch. All of these ions are typical for subluminous SNe Ia except for the \Crii\ in the second and third epochs, which mainly manifests itself as an absorption at 4700~\AA\ and possibly also a feature at $\sim$7000 \AA. The evidence for \Crii\ is weak, with \TIii\ doing most of the fitting in these regions but it does slightly improve the fits of the later two epochs.

\cite{fol10} found evidence of \Cii\ in the photospheric, optical phase spectra of SN 2006bt, which could be explained by circumstellar interaction with C-rich material. They suggest that this could indicate a preferred viewing angle along the plane of the accretion disk or it could be due to dense `blobs' of C separated from the rest of the ejecta, that when not viewed along the ejecta direction could result in lower line velocities, as were observed for this SN. To test for the presence of \Cii\ in the spectra of PTF10ops, it was included in the \textsc{SYNAPPS} fits but we find no evidence that it improves the fits so it was excluded. Similarly we checked for the presence of \Scii, which was found to be present in the photospheric optical spectra of the subluminous and peculiar SN Ia, PTF09dav \citep{sul11} but we also find no evidence for it in the spectra of PTF10ops.

\subsection{Spectral measurements}
\label{spec_measure}

The velocity, as measured from the \SiII\ 6355~\AA\ line is $\sim$10000 km s$^{-1}$ at $-0.6$ d. By comparing to fig.~19 of \cite{tau08}, the velocity of PTF10ops is found to be at the border between the typical velocities of subluminous 91bg-like objects and those of the low velocity gradient group as defined by \cite{ben05}. The \SiII\ line appears asymmetric in the spectra of PTF10ops after +10 d, with a second feature seen to the red side of the absorption feature. 

\begin{figure}
\includegraphics[width=8.8cm]{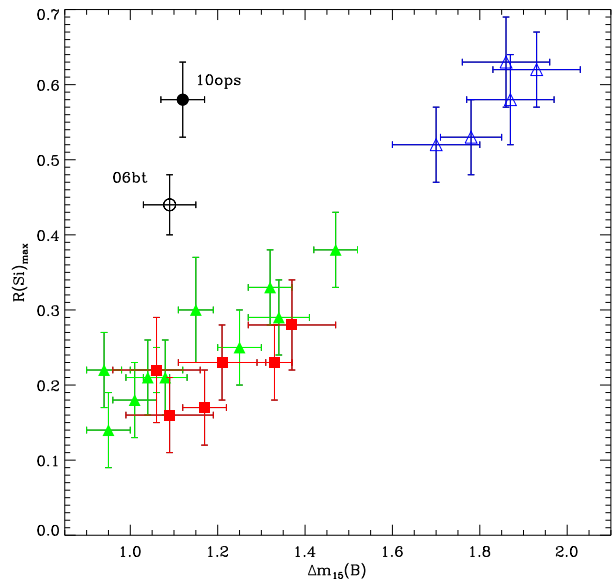}
\caption{Silicon ratio at maximum, $\mathcal{R}$(Si) against $\Delta$M$_{15}$(\textit{B}). The faint group is displayed as blue open triangles, the low velocity group as green solid triangles and the high velocity group as red solid squares. PTF10ops (solid black circle) and SN 2006bt (open black circle) are seen to be clear outliers from the pre-defined groups of \protect \cite{ben05}.}
\label{rsi_comp}
\end{figure}

Several methods have been investigated to sub-classify SNe Ia based on the relative fluxes and velocities of features present in their spectra. The ratio of the depth of the \SiII\ 5972~\AA\ to the \SiII\ 6355~\AA\ absorption features, $\mathcal{R}$(Si) at maximum compared to the $\Delta$M$_{15}$(\textit{B}) of the SNe, can be used to classify SNe Ia into distinct classes \citep{nug95}. Using the method in \cite{nug95}, the $\mathcal{R}$(Si) value of PTF10ops is calculated to be 0.58$\pm$0.05 and is compared to other SNe Ia in Figure \ref{rsi_comp}. PTF10ops is clearly seen to be an outlier in this plot of $\mathcal{R}$(Si) versus $\Delta$M$_{15}$(\textit{B}) with its small value of $\Delta$M$_{15}$(\textit{B}) and large value of $\mathcal{R}$(Si), which is thought to be indicative of a cool photosphere.

\begin{figure}
\includegraphics[width=8.8cm]{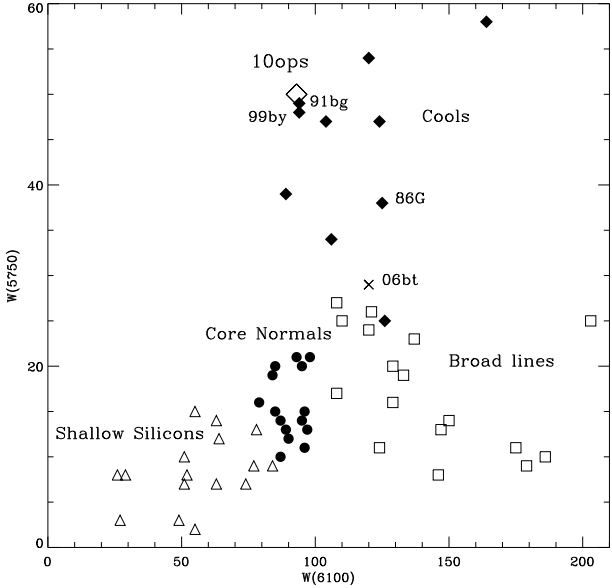}
\caption{The equivalent widths of the Si II features, W(6100) and W(5750), for a sample of SNe Ia as detailed in \protect \cite{bra06, bra09} and now including PTF10ops (open diamond) and SN 2006bt (cross). PTF10ops sits in the Cool subclass of SNe Ia as defined by \protect \cite{ben05}.}
\label{ew_comp}
\end{figure}

\cite{bra06,bra09} studied the spectra of a large sample of SNe Ia and based on certain spectral features, they defined four distinct subclasses of SNe Ia: core normal (CN), broad line (BL), cool (CL) and shallow silicon (SS). These classes are similar to those defined by \cite{ben05} as low-velocity gradient, high-velocity gradient and faint. Apart from a few SNe at the border between the groups, the low-velocity gradient group includes CNs and SSs, the high velocity gradient group is equivalent to the BL group and the faint class is the CLs. The equivalent widths of the Si II features 6355 \AA, W(6100) and 5972 \AA, W(5750) for PTF10ops are measured and found to have values of W(6100) = 93~\AA\ and W(5750) = 50 \AA. In Figure \ref{ew_comp}, the equivalent width relation of the sample of \cite{bra09} is compared to that of PTF10ops. PTF10ops is found to sit in the CL group and has values most similar to those of 1999by and 1991bg. 

\section{Supernova environment}
\label{sn_environ}

If PTF10ops is a SN Ia, as is suggested by its typical subluminous SNe Ia spectral features, then there are two scenarios for its origin; it is either located in the halo of the nearby spiral galaxy (maybe specifically in a globular cluster) or it has a very faint host galaxy. There is a small possibility that PTF10ops is the result of some exotic core-collapse explosion but we show that the likelihood of a massive star being ejected from the nearby spiral galaxy is very low and that a massive star formed at the position of the SN is not expected due to undetected star formation at this position.

Generally, the host galaxies of subluminous SNe Ia have been found to be massive E/S0 galaxies and hence, the events likely originate from old stellar populations. Recently a number of SN events have been discovered at distances of $\sim$23--41 kpc from the centre of their host galaxies e.g. PTF09dav \citep{sul11}, SN 2005E \citep{per10} and SN 2006bt \citep{fol10}. PTF09dav was a very subluminous SN Ia, SN 2005E was a Ca-rich SN and SN 2006bt was a SN Ia with a slowly declining lightcurve but spectral features typical of subluminous SNe Ia. The host of PTF09dav was a star-forming spiral galaxy while the hosts of SN 2005E and SN 2006bt were both older S0/a galaxies. 

If PTF10ops is associated with the massive spiral galaxy shown in Figure \ref{galaxy}, then at a distance of 148 kpc from the centre of its host, it is the most remote SN discovered to date. The halos of massive galaxies can easily extend to these types of distance as is the case for M31 which has been shown to have a metal-poor halo extending to a radius of 160 kpc \citep{kal06}. M31 has a stellar mass of log(M/\msol) = 11.0 \citep{ham07}, which is smaller than the value of (M/\msol) = 11.18$^{+0.04}_{-0.25}$ determined for the host galaxy of PTF10ops (see Section \ref{host_galaxy}). This suggests that the halo of the host of PTF10ops should extend to an even greater distance than that of M31. The halos of galaxies contain old stars, produced when the galaxy was formed, which suggests an old, metal-poor stellar population could be responsible for the progenitor system of PTF10ops. 

No galaxy is detected at the position of PTF10ops down to an absolute magnitude limit of $-12.0$ mag, which places a significant limit on the magnitude and size of any potential host galaxy. Before the discovery of PTF10ops, Type Ia SN 1999aw, a SN 1999aa-like object (high stretch, luminous SN) had the faintest detected host galaxy with an absolute magnitude of $-12.4$ \citep{str02}.  \cite{sul10} studied the host galaxies of SN Ia and found that in general, brighter, higher stretch SNe are found more frequently in low luminosity hosts, which PTF10ops is not and adds more weight to the scenario that PTF10ops is associated with the massive spiral galaxy. 

Faint, low metallicity halo galaxies have been discovered out to a projected distance of $\sim$150 kpc from the centre of M31 \citep{mar06, ric11}. These galaxies have magnitude in the range, $-10.2<M_V\leq-6.4$ and would not be detected in our deep image. There is also the possibility that PTF10ops exploded in an undetected globular cluster -- globular clusters have typical magnitudes of M$_V$ = $-7.0$ \citep{van91}. \cite{pfa09} studied the rate of SNe Ia in globular clusters and suggested that the dense stellar environments in a globular cluster might result in exotic SNe Ia. 

f the origin of PTF10ops was instead some sort of exotic explosion of a massive star and this spiral galaxy is the host, then it was most likely formed in either the inner disk or in a circumnuclear star cluster and ejected by an interaction with a massive galactic nucleus black hole. If the progenitor of PTF10ops was an 8 \msun\ star (the lower mass limit for a core-collapse SN progenitor), it would have to travel on average, at a velocity of $\sim$3600 km s$^{-1}$ to reach this position in its lifetime of $<40$ Myr, an extremely high velocity.  \cite{per10} details how the number of high-velocity stars detected at large distance from the galactic centre is small, only $\sim$20 hypervelocity stars have been discovered with velocities of 300-900 km s$^{-1}$ at distances of 20--120 kpc from the Milky Way centre.  Therefore, the likelihood of a massive star progenitor at this distance from the galaxy is very low. Combining this with the SN Ia spectral and lightcurve properties, the likelihood of a massive star as the progenitor of PTF10ops can be ruled out. 

Ruling out a massive star origin, regardless of whether PTF10ops exploded in the halo of the massive galaxy, a halo galaxy or an associated globular cluster, these environments suggest an old (possibly very old) WD population with low metallicity as the most likely progenitor scenario for PTF10ops, which will be discussed in more detail in Section \ref{results}.

\section{Discussion and Conclusions}
\label{results}

\begin{table}
 \caption{Best match for the properties of PTF10ops to normal and subluminous SNe.}
 \label{table_comp2}
 \begin{tabular}{@{}lccccccccccccccccccccccccccccccc}
  \hline
  \hline
  & Normal & Subluminous \\
  \hline
Best template fit& & Yes\\
Absolute magnitude &&Yes\\
Lightcurve stretch &Yes&\\
(\bv)$_{max}$&&Yes$^a$\\
$\Delta$M$_{15}$(\textit{B})&Yes&\\
Rise time&Yes&\\
$I$ band profile shape&&Yes\\
\nick\ mass &&Yes\\
Spectral features present &&Yes\\
Velocity of spectral features &&Yes$^a$\\
$\mathcal{R}$(Si)&&Yes\\
\SiII\ EW ratio&&Yes\\
\hline
 \end{tabular}
 \begin{flushleft}
 $^a$The value is intermediate between normal and subluminous events but closest to that of a subluminous SN Ia.\\
  \end{flushleft}
\end{table}

Table \ref{table_comp2} shows the best matches out of normal and subluminous SNe Ia for the properties of PTF10ops. The long rise-time and normal-width lightcurve suggest a normal SN Ia. However, its absolute magnitude is more like that of a subluminous SN Ia and it lacks secondary maxima in its $r$ and $i$ band lightcurves, which is thought to be caused by a low \nick\ mass and/or a high value of mixing in the ejecta. A low \nick\ mass of 0.17$\pm$0.01 \msun\ is estimated using a comparison with the rise-time of the subluminous template used in the lightcurve fit. The features in the optical spectra of PTF10ops are also at odds with its stretch value; the spectral features include those of \TIii, which are typical of the cool photospheres of subluminous SNe Ia. 

Its first three photospheric phase spectra were analysed using the spectral fitting code, \textsc{synapps} and can be explained using features that are characteristic of subluminous SNe Ia, and it is not necessary to include uncommon elements such as \Cii, seen in SN 2006bt and \Scii, seen in PTF09dav to explain the spectra of PTF10ops. The velocities of the spectral features were measured to be similar to those of other subluminous SNe and the calculated value of $\mathcal{R}$(Si) is most similar to that of the faint group of \cite{ben05}. Thus with its slowly declining lightcurve, it falls well outside the established groups on a plot of $\mathcal{R}$(Si) against $\Delta$M$_{15}$(\textit{B}). Measurements of the equivalent widths of the Si II features 6355 \AA,W(6100) and 5972 \AA, W(5750) for PTF10ops show that the values are most similar to those of subluminous SNe. This suggests that PTF10ops had a cool photosphere, which is in agreement with the \TIii\ features that are formed at lower temperatures and hence are seen in cool, subluminous SNe Ia. 

No host galaxy was found at the position of PTF10ops down to a limiting magnitude of $r\geq-12.0$ mag. However, there is a massive spiral galaxy located at a separation of $\sim$148kpc from the SN position, which was determined to have the same redshift as was estimated from the spectral fits of PTF10ops. See Section \ref{sn_environ}, for more details on the properties of the environment of PTF10ops.

An important question about these unusual objects such as PTF10ops, PTF09dav, SN 2006bt and SN 2005E, is why do they appear to show a preference for large distances from their host galaxies and/or potentially very faint host galaxies? Since similar, unusual SNe have not been observed in the `normal' environments of SNe Ia, it is likely that environmental effects, such as age and metallicity could be very important in producing these events. Metallicity may play a role in the explosion of these SNe, but if these SNe were formed due to low metallicity, then we would have expected to find them in other low metallicity environments, such as in dwarf galaxies. However, none have yet been found. Therefore, we conclude that while metallicity may influence the progenitor evolution, it is not likely to be the dominating factor. 

The most fundamental property of these remote sites is instead that they are likely to host very old stellar populations ($>10$ Gyr). If PTF10ops formed in the halo of this massive galaxy, then it could have an age equivalent to its redshift of $\sim$13 Gyr, and hence, a \textit{very} old stellar population may be present, which could lead to a different SN explosion than is generally seen for SNe Ia. However, the exact progenitor channel and mechanism is still unclear. A globular cluster origin for PTF10ops is also a possibility with the dense stellar environments of globular clusters being able to produce unusual, exotic SNe. However, globular clusters are found throughout the halos of galaxies, not just at large distances from the centre so either the very old stellar populations or very low metallicity or some combination of the two must also play a role.

What kind of progenitor star would lead to a SN like PTF10ops, can not be easily understood. It is difficult to explain the diversity of subluminous SN Ia events
with one single progenitor scenario, and several different channels
may play a role. One possible scenario for the production of a subluminous SN Ia with a normal-width lightcurve has been suggested by \cite{pak10}. They modelled the mergers of two equal-mass WD with masses in the range 0.83--0.9 \msun\ and found that they could lead to a subluminous SN Ia explosion. 
However, some of the properties of subluminous 1991bg-like events are not recreated in the models. These discrepancies include that the $\Delta$m$_{15}$(\textit{B}) are lower (1.4--1.7) than for 1991bg-like objects i.e. that the lightcurve is too broad. Although the $\Delta$m$_{15}$(\textit{B}) of PTF10ops was smaller than this range, a merger of two equal-mass WD could produce SNe that have features that fall in between those of normal and 
subluminous events, such as PTF10ops. These could have many of the features of a subluminous SN such as a low \nick\ mass, no secondary maxima in its lightcurve in the redder filters but a broader lightcurve with a longer rise-time. The main discrepancy so far is that the models predict a redder colour around maximum light compared to PTF10ops but further work is ongoing to investigate these issues. The importance of the initial conditions on the mass transfer phase is also highlighted in \cite{dan11}; a very dynamical merger is needed so that the compression is sufficient to cause a detonation. A more direct collision could speculatively occur in the core of a globular cluster where the conditions would be more suitable for producing the compression need to cause a detonation \citep{ros09}. 

Previous SN searches have been biased toward observing massive galaxies and it is only with current, untargeted searches that we are starting to detect more SNe at positions far from massive galaxies, which will hopefully allow us to build up a meaningful sample of these unusual types of event to better understand their origins and explosion mechanisms. Understanding why these objects exist with normal stretch lightcurves but very different spectral properties to normal SNe Ia, may also help in our understanding of the bulk of the SN Ia population. These unusual events such as PTF10ops and SN 2006bt should not bias SN Ia cosmological samples as long as spectra are obtained to identify their subluminous features, regardless of their lightcurve properties.

\section{Acknowledgements}

MS acknowledges support from the Royal Society. AG and MS acknowledge support from the Weizmann-UK "making connection" program.  The Weizmann Institute PTF partnership is funded in part by the Israeli 
Science Foundation (ISF) via a grant to AG. Joint WIS-Caltech activity 
is funded by a Binational Science Foundation (BSF) grant to AG and SRK.
AG further acknowledges support from the EU/FP7 via a Marie Curie IRG
fellowship and an ARCHES prize from the German BMBF. This work was supported by the Science and Technology Facilities Council. EOO is supported by NASA grants. EOO and DP are both supported by
Einstein fellowships. S.B.C. acknowledges generous financial assistance from Gary \& Cynthia Bengier, the Richard \& Rhoda Goldman Fund, NASA/{\it Swift} grants NNX10AI21G and GO-7100028, the TABASGO Foundation, and NSF grant AST-0908886. This publication has been made possible by the participation of more than 10,000 volunteers in the Galaxy Zoo Supernovae project, http://supernova.galaxyzoo.org/authors.

The WHT is operated on the island of La Palma by the Isaac Newton Group in the Spanish Observatorio del Roque de los Muchachos of the Instituto de Astrof\'isica de Canarias. The Liverpool Telescope is operated on the island of La Palma by Liverpool John Moores University in the Spanish Observatorio del Roque de los Muchachos of the Instituto de Astrofisica de Canarias with financial support from the UK Science and Technology Facilities Council. Observations were obtained with the Samuel Oschin Telescope at the Palomar Observatory as part of the Palomar Transient factory project, a scientific collaboration between the California Institute of Technology, Columbia Unversity, La Cumbres Observatory, the Lawrence Berkeley National Laboratory, the National Energy Research Scientific Computing Center, the University of Oxford , and the Weizmann Institute of Science. Some of the data were obtained with the W.~M.~Keck Observatory, which is operated as a scientific partnership among the California Institute of Technology, the University of California and the National Aeronautics and Space Administration. The observatory was made possible by the generous financial support of the W.~M.~Keck Foundation. 

SNIFS on the UH 2.2-m telescope is part of the Nearby Supernova Factory II project,
a scientific collaboration between the Centre de Recherche Astronomique
de Lyon, Institut de Physique Nucl\'eaire de Lyon, Laboratoire de Physique
Nucl\'eaire et des Hautes Energies, Lawrence Berkeley National Laboratory,
Yale University, University of Bonn, Max Planck Institute for Astrophysics,
Tsinghua Center for Astrophysics, and the Centre de Physique des Particules
de Marseille. Based on observations made with the NASA/ESA Hubble Space Telescope, obtained from the data archive at the Space Telescope Science Institute. STScI is operated by the Association of Universities for Research in Astronomy, Inc. under NASA contract NAS 5-26555."

This research has made use of the NASA/IPAC Extragalactic Database (NED) which is operated by the Jet Propulsion Laboratory, California Institute of Technology, under contract with the National Aeronautics and Space Administration.

\end{document}